\documentclass[aps,floats,floatfix,showpacs,twoside,preprint,superscriptaddress]{revtex4}

\usepackage{latexsym}
\usepackage{amsfonts}
\usepackage{graphicx}
\usepackage{amsbsy}
\usepackage{float}
\usepackage{epstopdf} 
\DeclareGraphicsExtensions{.pdf,.eps,.png,.jpg,.mps}

\begin{document}

\title{Scaling of Seismic Memory with Earthquake Size }
\author{Zeyu Zheng}
\affiliation{Department of Environmental Sciences, 
Tokyo University of Information Sciences, Chiba 265-8501, Japan}
\author{Kazuko Yamasaki}

\affiliation{Department of Environmental Sciences, 
Tokyo University of Information Sciences, Chiba 265-8501, Japan}
\affiliation{Center for Polymer Studies and Department of Physics, Boston
University, Boston, MA 02215, USA}

\author{Joel Tenenbaum}
\affiliation{Center for Polymer Studies and Department of Physics, Boston
University, Boston, MA 02215, USA}
\author{Boris Podobnik}

\affiliation{Center for Polymer Studies and Department of Physics, Boston
University, Boston, MA 02215, USA}
\affiliation{Faculty of Civil Engineering, University of Rijeka, Rijeka, Croatia}
\author{H.~Eugene~Stanley}

\affiliation{Center for Polymer Studies and Department of Physics, Boston
University, Boston, MA 02215, USA}

\begin{abstract}

It has been observed that the earthquake events possess short-term  memory, i.e.
that events occurring in a particular location are dependent on the
short history of that location.  We conduct an analysis to see whether
real-time earthquake data also possess long-term memory and, if so, whether such
autocorrelations depend on the size of earthquakes within close
spatiotemporal proximity.  We analyze the seismic waveform database
recorded by 64 stations in Japan, including the 2011
 ``Great East Japan Earthquake'', 
 one of the five most powerful earthquakes ever recorded which resulted in 
a tsunami and devastating nuclear accidents.
 We explore the question of seismic 
memory through use of mean conditional intervals and detrended
fluctuation analysis (DFA).  
We find that the waveform sign series show long-range
 power-law anticorrelations
while the interval series show long-range power-law correlations. We find
 size-dependence in  earthquake auto-correlations---as earthquake size  
 increases, both of these correlation behaviors strengthen. We also find that the DFA
scaling exponent $\alpha$ has no dependence on earthquake hypocenter depth or
epicentral distance.
\end{abstract}

\maketitle
\section{Introduction}
Many complex physical systems exhibit complex dynamics in which subunits
of the system interact at widely varying scales of time and space \cite{MFS,Bunde}. These
complex interactions often generate very noisy output signals which
still exhibit scale-invariant structure.  Such complex systems span
areas studied in physiology \cite{Ashk20}, finance \cite{Engle82}, and
seismology 
\cite{Bak,Corral04,Corral05,Lippiello07,Lippiello08,Bottiglieri,Kagan,Lennartz10,Livina05,Lennartz08}.

In seismology the study of seismic waves is both scientifically
interesting and of practical concern, particularly in such applied areas
as engineering.  A better understanding of seismic waves is immediately
applicable in the design of structures for earthquake-prone areas
\cite{Mayeda,Hisada,Hisada2}. It also allows scientists to better
understand the underlying mechanisms that drive earthquakes
\cite{Bensen,Xu2,Okada,Ide,Shapiro,Campillo}.  In seismology, temporal
and spatial clustering are considered important properties of seismic
occurrences and, together with the Omori law (dictating aftershock
timing) and the Gutenberg-Richter law (specifying the distribution of
earthquake size), comprise the main starting requirements to be
fulfilled in any reasonable seismic probabilistic model. Analyzing the
timing of individual earthquakes, Ref.~\cite{Bak} introduces the scaling
concept to statistical seismology.  The recurrence times are defined as
the time intervals between consecutive events, $\tau_i = t_i -
t_{i-1}$. In the case of stationary seismicity, the probability density
$P(\tau)$ of the occurrence times was found to follow a universal
scaling law
\begin{equation}
 P(\tau) = R f(R \tau) 
\end{equation}
where $f$ is a scaling function and $R$ is the rate of seismic
occurrence, defined as the mean number of events with $M \ge M_c$
\cite{Corral04}. Reference~\cite{Corral05,Lippiello07} has demonstrated
how the structure of seismic occurrence in time and magnitude can be
treated within the framework of critical phenomena.

Recently, a few papers have analyzed the existence of correlations
between magnitudes of subsequent earthquakes
\cite{Corral05,Lippiello07}.  Analyzing earthquakes with $\tau$ greater
than 30 minutes, Ref.~\cite{Corral05} reported possible magnitude
correlations in the Southern California catalog. Magnitude correlations
have often been interpreted as a spurious effect due to so called
short-term aftershock incompleteness (STAI) \cite{Kagan}.  This
hypothesis assumes that some aftershocks, especially small events, are
not reported in the experimental catalogs, which is in agreement with
the standard approach that assumes interdependence of earthquake
magnitudes implying no memory in earthquakes.

However, recent work has also challenged this interpretation.
Reference~\cite{Lippiello08} reports the existence of magnitude
clustering in which earthquakes of a given magnitude are more likely to
occur close in time and space to other events of similar magnitude.
They find that a subsequent earthquake tends to have a magnitude similar
to but smaller than the previous earthquake.
Reference~\cite{Lippiello07} also reports the existence of magnitude
correlations and additionally demonstrates the structure of these
correlations and their relationship to $\Delta t$ and $\Delta r$, where
the latter represents the distance between subsequent epicenters.
Reference~\cite{Lennartz10} creates a model to explain these magnitude
correlations.  They note that the Omori law and ``background tectonic
cycles'' are responsible for clustering in interoccurrence times.
Additionally, Refs.~\cite{Livina05} and \cite{Lennartz08} find that the
distribution of recurrence times strongly depends on the previous
recurrence time such that small and large recurrence times tend to
cluster in time.  This dependence on the past is reflected in both the
conditional mean recurrence time and the conditional mean residual time
until the next earthquake.

Since it is our hypothesis that long-range autocorrelations exist in
seismic waves, we first note that long-range power-law autocorrelations
are quite common in a large number of natural phenomena ranging from
weather \cite{Yamasaki3,Gozolchiani,BP05}, and physiological systems
\cite{Lennartz,Ashk20,Kantelhardt2,Stanley3,Karasik}, to financial
markets \cite{Mantegna,Yamasaki,Wang2,Wang3,Stanley,BPPnas09,BPPnas10}.

In addition to analyzing the raw waveform, it is also common to analyze
related time series, such the time series generated by taking the sign
or magnitude of the waveform \cite{Ashk20}.  Reference~\cite{Ashk20}
reports an empirical approximate relation at small time scales for the
scaling exponents calculated for sign, magnitude, and the original time
series, $\alpha_{\rm sign} = 1/2 (\alpha_{\rm magnitude} + \alpha_{\rm
  original})$, in physiology. The study of magnitude and sign time
series is important in physiology because the magnitude time series
exhibits weaker autocorrelations and a scaling exponent closer to the
exponent of an uncorrelated series found when a subject is unhealthy
\cite{Ashk20}. Diagnostic power in physiology has been confirmed for
sign time series as well---the sign time series of heart failure
subjects exhibit scaling behavior similar to that observed in the
original time series, but significantly different that of healthy
subjects \cite{Ashk20}.  Understanding the correlation properties of
these three time series allows us to also understand the underlying
processes generating them.

Our investigation and discussion is organized as follows.  First, we
study the autocorrelations of interval series by using the mean
conditional technique. Second, we employ detrended fluctuation analysis
(DFA) \cite{CKP,Hu,Chen} and find  long-range power-law
 autocorrelations in the sign and interval time series.  For
the interval time series we find a positive regression between the DFA
scaling exponent $\alpha$ and earthquake size (measured by the Richter
magnitude scale $M$ or seismic moment $M_0$), while for the sign time
series we find an inverted regression between $\alpha$ and earthquake
magnitude. Thus we report that the observed autocorrelation depends on
earthquake size, both in the sign and interval time series.  We also find
that the scaling exponent $\alpha$ has no dependence on hypocenter depth
or epicentral distance.

\section{Data}

Seismic waves are unique in that they have non-stationarities of a much
larger order than those of any other known natural signal.  Large
earthquakes are characterized by a maximum amplitude that is often
$>100$ times larger than the mean amplitude [see Fig.~1(a)]. This is a
limitation that makes seismic waves difficult to analyze using
traditional analysis. Although we might want to use detrended
fluctuation analysis (DFA) \cite{CKP,Hu,Chen,Chen2}, originally proposed to
study the correlations in a time series in the presence of
non-stationarities commonly observed in natural phenomena, the level of
non-stationarity in earthquakes is so large that DFA is inappropriate
regardless of the order of the polynomial fit applied \cite{Chen}. Thus,
due to lack of methods for highly non-stationary signals, we do not
analyze correlations in the series of magnitudes, but instead analyze
the correlations in the sign series [Fig.~1(c)] and interval series
[Fig.~1(d)].  For our data, we use the seismic waveform database from
the National Research Institute for Earth Science and Disaster
Prevention (NIED) F-net (Full Range Seismograph Network of Japan), which
records continuous seismic waveform data $w_t$ by using broadband sensors in
64 stations in Japan [see Fig.~1(a)]. 
In our study we select 46 stations (ADM, AOG, ASI, HID, HJO, HRO, IGK, IMG, INN, IYG, IZH, KGM, KMU, KNM, KNP, KNY, KSK, KSN, KSR, KYK, MMA, NKG, NOK, NOP, NRW, NSK, OSW, SAG, SHR, SIB, TAS, TGA, TGW, TKO, TMC, TSA, TYM, TYS, UMJ, WTR, YAS, YNG, YSI , YTY, YZK, ZMM), based on locations and integrity of data series.
Seismic signals are recorded in three directions: (1) U (up-down
with up positive), N (north-south with north positive), and E (east-west
with east positive) \cite{Okada}.  In this paper, we report results from
the vertical dimension only (U data), since the results for the
horizontal data (N and E) data are very similar.   Sampling intervals
have five recording frequencies: 80Hz, 20Hz, 1Hz, 0.1Hz, and 0.01Hz.  We
study earthquake coda wave data with 1Hz sampling interval for the
year 2003, together with selected earthquake coda wave data from 11
March 2011.  We note that, because of the interaction between
earthquakes, not all earthquakes can be employed in our analysis (see
Appendix A).  The data from 11 March 2011 is selected because it
contains the notable 2011 Tohoku earthquake (``Great East Japan
Earthquake'') which resulted in the tsunami that caused a number of nuclear accidents. 
  We also add two large earthquakes ($M=7.3$ and $M=7.6$) to
our study, which also occurred the same day as aftershocks. 

We employ the following procedure to create our time series:
  \begin{itemize}
 \item[{(i)}]
 For  each selected earthquake (see Appendix A)
 we create a new time series, the normalized waveform denoted by $w_t$
 out of the raw seismic acceleration waveform data
\begin{equation}
  w_{norm}\equiv(w_t -\overline {w}) / \sqrt{\overline{w_t^2} -\overline{w}^2}.
\end{equation}
 \item[{(ii)}]
  From the time series $w_{norm}$  we define a new sub-series   
 $w_t'$,  starting at time coordinate where    
 maximum  $w_t$ occurs and terminating 
  at the end of the normalized waveform $w_t'$ (see inset in  Fig1(a)). 
\item[{(iii)}]
 Let the time series $t_i$ denote the points in time when $w_t'$ changes 
sign, with $t_i  < t_{i+1}$.
 We define (see Fig 1(c)) the interval series by 
\begin{equation}
 \tau_i \equiv t_i -  t_{i-1}. 
 \label{tau}
 \end{equation}
\item[{(iv)}]
 The sign series (see Fig. 1(d)) is defined by 
 \begin{equation}
 s_t \equiv sgn(w_t') 
  \label{sign}
 \end{equation}
  \end{itemize}

Note that our definition of interval is different than 
 that recently defined in  several papers,  where
the return intervals $\tau$   have studied 
 between consecutive  fluctuations above a volatility
threshold $q$ in different complex systems.
 The probability density function (pdf) of return
intervals $P_q(\tau)$ scales with the mean return interval 
 as 
\begin{equation}
  P_q(\tau) =  \overline {\tau}^{-1}  f(\tau / \overline {\tau} )
\end{equation}
 where $f()$ is a stretched exponential  \cite{Yamasaki,Wang2,Wang3}. 
 Since, on average, there is one volatility above the threshold $q$ for every
$\overline{\tau}_q$ volatilities, then it holds that \cite{BPPnas09}   
\begin{equation}
1/\overline{\tau}_q\approx \int_q^\infty P(|R|)d|R|=P(|R|>q)\sim q^{-\alpha}.
\label{PR}
\end{equation}
 For the time intervals $\tau_q$  between events given by 
 fluctuations $R$ where 
  $R > q$
 Ref.~\cite{BPPnas09}  derived  that $ \overline {\tau_q} $, the average of  $ \tau_q $,
 obeys a scaling law,
\begin{equation}
   \overline {\tau_q} = q^{\alpha} 
   \label{scale1}
\end{equation}
  where by $\alpha$ denotes our estimate of the tail exponent
   probability density function, $P(|R|^{1 + \alpha})$.  
 Similarly, if $P(|R|)$ follows an exponential function 
 $P(|R|) \propto \exp(- \beta |R|)$, then employing  Eq.~(\ref{PR}) we easily
  derive 
  \begin{equation}
 \overline{\tau}_q \propto  \exp{(\beta q)}.
\label{scale2}
\end{equation}
   Eq.~(\ref{scale2}) can be used as a new method for estimation of the 
   exponential parameter $\beta$.

   \section{ Memory of interval time series}

 Returning to waveform data,  we begin
analyzing the series by studying the conditional mean
\begin{equation}
\langle \tau | \tau_0 \rangle /  \overline \tau 
\end{equation}
which gives the mean value of $\tau$ (see Eq.~(\ref{tau})) immediately following a given term
$\tau_0$, normalized in units of $\overline \tau$.  The conditional mean
gives evidence of whether seismic memory exists in the intervals in the
form of correlations or anticorrelations.  For example, should
correlations exist, one would expect the mean interval to be shorter in
the window immediately following a small interval.

Indeed, Fig.~2 shows that the large intervals $\tau$ tend to follow
large initial $\tau_0$ and small $\tau$ follow small $\tau_0$ indicating
the existence of (positive) correlations in the interval time series.
We also note that the autocorrelations tend to be stronger for the
subset associated with larger earthquakes than for those associated with
smaller earthquakes.

To expand on this we also extend our investigation to longer range
effects.  We investigate the mean interval after a cluster of $n$
consecutive intervals that are either entirely above the series mean or
entirely below it.  We denote clusters that are entirely above the
series mean with a ``$+$'' and clusters below the series mean with a
``$-$''.  Fig.~3 shows the mean interval $\tau$ that follows a
$\tau_0(n)$ defined as a cluster size of $n$.  We find that for ``$+$''
clusters---shown by open symbols---the mean interval increases with the
size of the cluster $n$. This is the opposite of what we find for
``$-$'' clusters---shown as closed symbols.  The results indicate the
existence of at least short-term memory in the interval time
series. Furthermore, we find that the mean interval increases with the
seismic magnitude. However, this relationship breaks at the high end of
the Richter magnitude scale $M>6.5$.

\section{Detrended fluctuation analysis}

Many physical, physiological, biological, and social systems are
characterized by complex interactions between a large number of
individual components, which manifest in scale-invariant
correlations \cite{MFS,Bunde,Tak,Kob82}.  Since the resulting observable
at each moment is the product of a magnitude and a sign, many recent
investigations have focused on the study of correlations in magnitude
and sign time series
\cite{Ashk20,Kantelhardt2,Kant2002,Hu,Plamen04itt,Livina,Engle82}.  For
example, the time series of changes $\delta \tau_{i}$ of heartbeat
intervals \cite{Ashk20,Kant2002,Kantelhardt2}, physical activity levels
\cite{Hu}, intratrading times in the stock market \cite{Plamen04itt},
and river flux values \cite{Livina} all exhibit power-law
anticorrelations, while their magnitudes $|\delta \tau_{i}|$ are
positively correlated.  A common means of finding autocorrelations
hidden within a noisy non-stationary time series is detrended
fluctuation analysis (DFA)\cite{CKP,Hu,Chen}.  In the DFA method, the
time series is partitioned into pieces of equal size $n$.  For each
piece, the local trend is subtracted and the resulting standard
deviation over the entire series is obtained.  In general, the standard
deviation $F(n)$ of the detrended fluctuations depends on $n$, with
smaller $n$ resulting in trends that more closely match the data.  The
dependence of $F$ on $n$ can generally be represented as a power law
such that
\begin{equation}
  F(n) \propto n^\alpha,
\end{equation}
where $\alpha$ is the scaling exponent---sometimes referred to as the
Hurst exponent---to be obtained empirically.  DFA therefore can
conceptually be understood as characterizing the motion of a random
walker whose steps are given by the time series.  $F(n)$ gives the
walker's deviation from the local trend as a function of the trend
window.  Because the root mean square displacement of a walker with no
correlations between his steps scales like $\sqrt(n)$, we can expect a
time series with no autocorrelations to yield an $\alpha$ of 0.5.
Similarly, long-range power-law correlations in the signal (i.e. large
terms follow large terms and small terms follow small terms) manifest as
$\alpha > 0.5$.  Power-law anticorrelations within a signal will result
in $\alpha < 0.5$.  Additionally, DFA can be related to the
autocorrelation as follows: if the autocorrelation function $C(L)$ can
be approximated by a power law with exponent $\gamma$ such that
\begin{equation}
 C(L) \propto L^{-\gamma},
\end{equation}
then $\gamma$ is related to $\alpha$ by \cite{CKP}
\begin{equation}
 \alpha \approx 1 - \gamma/2.
\end{equation}

Another reason we employ the DFA method is that it is appropriate for
sign time series \cite{Kantelhardt2}. Other techniques for the detection
of correlations in non-stationary time series are not appropriate for
sign time series. Also, because the sign and interval time series have
affine relations, the analysis of sign will be helpful in understanding
the intervals.  However, the DFA gives biased estimates for the
power-law exponent in analysis of anticorrelated series \cite{Hu}, and
so in order to improve the accuracy of analysis, we integrate the time
series before we employ the standard DFA procedure. 

For the 2011 Tohoku earthquake, also known as the ``Great East Japan Earthquake'',
 we present the fluctuation function $F(n)$ of the coda 
 wave, measured at KSN station, as typical
examples of sign and intervals time series (Fig.~4).
 By using DFA, we find, for
most coda waves after earthquakes, that the time series of the
intervals  are consistent  with  a power-law correlated behavior 
$\alpha =0.69 $, while the 
sign time series of Eq.~(\ref{sign}) are consistent with a 
 power-law anti-correlated 
 behavior ($\alpha =0.32$).  
 The results therefore indicate
that for the interval series large increments are more likely to be
followed by large increments and small increments by small increments.
These results are in agreement with the results of the correlation
analysis reported in Section 3.  In contrast, anticorrelations in the
sign time series indicate that positive increments are more likely to be
followed by negative increments and vice versa.

For the entire set of sign time series comprising our sample we calculate  
 the average DFA scaling exponent  
   $\overline {\alpha} =0.34 \pm 0.09$ 
 indicating anticorrelations, and for the  interval time series
   we calculate the average DFA scaling exponent 
   $\overline {\alpha} =0.58 \pm 0.08$ 
   indicating correlations.
    For the different stations measuring the 2011 Tohoku earthquake
    we find that for the sign time series, $\overline {\alpha} =0.29 \pm 0.05$ 
 and for the  interval time series, $\overline {\alpha} =0.66 \pm 0.07$. 

\section{Relation between earthquake moments and scaling exponents of sign and interval series} 
Because large earthquake events release such extraordinary amounts of
energy, it is reasonable to ask whether their occurrence influences
local wave dynamics.  To this end, we study interval time series of
coda waves from earthquakes occurring in 2003, also including the
particularly large events of 11 March 2011, when three events $M>7$
occurred in the same day.  Fig.~5(a) shows the DFA scaling exponent of
the sign series versus seismic moment, where seismic moment is a
quantity used to measure the size of an earthquake.  We find a
decreasing functional dependence between the DFA exponent of the sign
series and the seismic moment of the proximal earthquake with slope
$\gamma = -0.028 \pm 0.002$, indicating that the DFA exponent 
decreases approximately with
seismic moment.  Note that because most of the exponents are $< 0.5$,
this indicates the presence of ever stronger anticorrelations in the
time series as earthquake magnitude increases.  Note, however, that the data
 break with this trend for very large
earthquakes (Richter magnitude scale $> 6.6$ or seismic moment $>
10^{19}$). 

We also find similar results in the interval series, the
difference being that the anticorrelations become correlations.
Fig.~5(b) shows that the DFA interval exponent and seismic moment
exhibit a {\it positive\/} functional dependence with slope
 $\gamma = 0.025 \pm 0.002$ so
that the DFA exponent increases with increasing seismic moment.  Because
most of the exponents for the interval series are $>0.5$, this indicates
that the series show stronger correlations for increasing seismic
moment.  Again, as with the sign series, we find a deviation from this
trend for very large earthquakes.

Having observed the influence of seismic moment on autocorrelations, we
now investigate whether other readily observable factors such hypocenter
depth and epicentral distance (the distance from the event to the
recording station) also contribute.  Specifically, we would like to
explore whether there is evidence that such long-term memory is affected
by the spreading process as seismic waves disseminate outward from their
epicenter to a recording station or whether the memory observed is
strictly due to the seismic activity.  Fig.~6 shows that the DFA
exponent for both interval and sign series are independent of both
hypocenter depth and epicentral distance.  From these results we
speculate that the DFA exponent is mainly a result of the
characteristics of the hypocenter rather than the process by which the
seismic waves are spread.

For moderately large earthquakes ($M_0=10^{14}\sim10^{19}$), we approximate
 the relation between the DFA scaling exponent and seismic
moment through the empirical formula

\begin{equation} 
      \alpha \approx a ~ log_{10}(M_0) +c  
\end{equation}

where $a=-0.028$, $c=0.797$ for the sign time series and where $a=0.025$,
$c=0.174$ for the interval time series. Since

\begin{equation}
      M = (log(M_0) - 9.1)/1.5,
\end{equation}

we can also write
\begin{equation}
      \alpha  \approx a ( 1.5 M + 9.1) + c = a' ~M + c',
\end{equation}

where $a'=-0.042$, $c'=0.542$ for the sign series, 
and $a'=0.037$, $c'=0.398$ for the interval series.

We note that similar size dependence in Hurst exponent 
 was found in Ref.~\cite{Eisler} where 
 Hurst exponents of financial time series 
increase logarithmically with company size. 
\section{Summary}
We analyze seismic coda waves during earthquakes, finding long-range power-law
autocorrelations in both the interval and sign time series.  The sign 
series generally display power-law anticorrelated behavior, with 
 anticorrelations
becoming stronger with larger earthquake events, while the interval
series generally display power-law correlated behavior, 
 with correlations also
becoming stronger with larger earthquake events.  We also show that
while the DFA autocorrelation exponent is influenced by the size of the
earthquake seismic moment, it is unaffected by earthquake depth or
epicentral distance.  Our findings are in contrast with a standard
approach which assumes independence in earthquake signals and thus have
strong implications on the ongoing debate about earthquake
predictability \cite{SornettePNAS}.

\section{Acknowledgements}
We thank S. Havlin for his constructive suggestions, 
and thank JSPS for grant of "Research project for a
 sustainable development of economic and social structure 
 dependent on the environment of the eastern coast of Asia" 
 that made it possible to complete this study. 
 We also thank the National Science Foundation and the
Ministry of Science of Croatia for financial support.
\section{Appendix: The Selection of Earthquakes}
In some regions it is common for multiple earthquakes to occur in short
succession.  In many cases, because the interoccurrence times are so
short, the coda waves can be derived from more than one earthquake.
This is especially true for large earthquakes with many aftershocks
\cite{Utu}.  In order to make sure that the coda waves we study are
the effects of only one earthquake, we need a way of determining which
earthquakes are independent.  We use the following two functions to
determine the sphere of influence and duration of each earthquake by
using the Richter magnitude scale M \cite{Utu}.  We select only those
earthquakes that have no larger earthquake in their spatiotemporal
sphere of influence,
\begin{equation}
t\approx 10^{(M-4.71)/1.67} 
\end{equation}
and
\begin{equation}
 R\approx 2 \times 10^{(M+1)/2.7},
\end{equation}	
where $t$ is the duration and $R$ is the sphere radius of influence.
The two functions are empirical formulas based on an analysis of
earthquakes in Japan \cite{Utu}. The $10^{M+1.0}/2.7$ is an empirical
formula that indicates the maximum radius that a human can feel an
earthquake, especially for the earthquakes in Japan.

\begin{figure}[b]
\centering \includegraphics[width=0.75\textwidth]{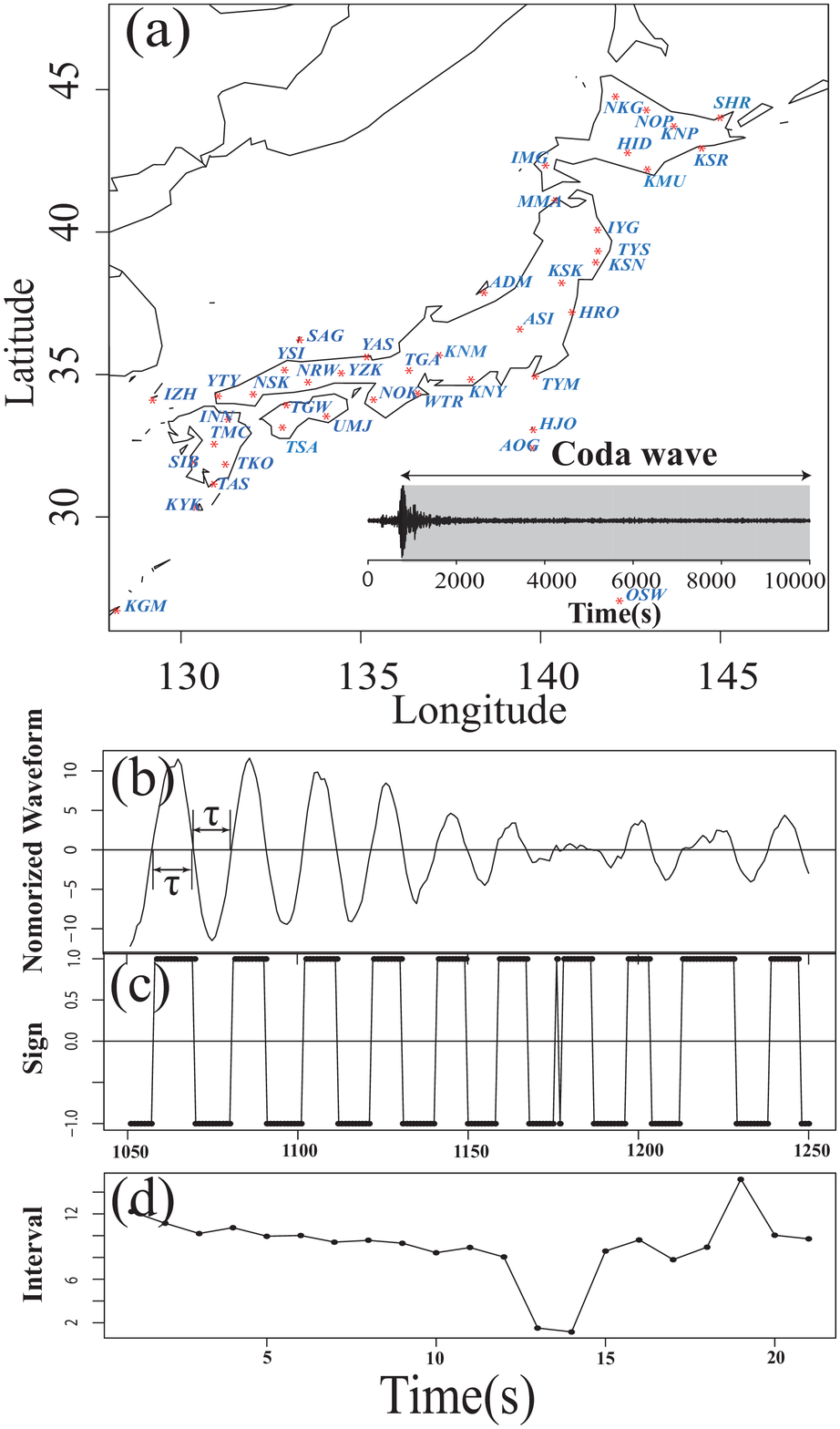}\\
\caption{(a) Location map for the 46 broadband stations of Full
 Range Seismograph Network of Japan (F-net) (red snow marks). Inset: 
 An example of a record of a seismic wave (Up-Down component). 
 (b) A part of the coda wave series indicated in inset of (a), as an example. 
 (c) An example sign time series where the positive sign (+1) represents a positive waveform, 
 and the negative sign (-1) represents a negative waveform in coda wave series 
 of seismic wave. (d) Interval time series ($\tau$) of the coda wave series 
 for a subset of the record shown in (b).
}
\label{1}
\end{figure}

\begin{figure}[b]

\centering \includegraphics[width=0.7\textwidth]{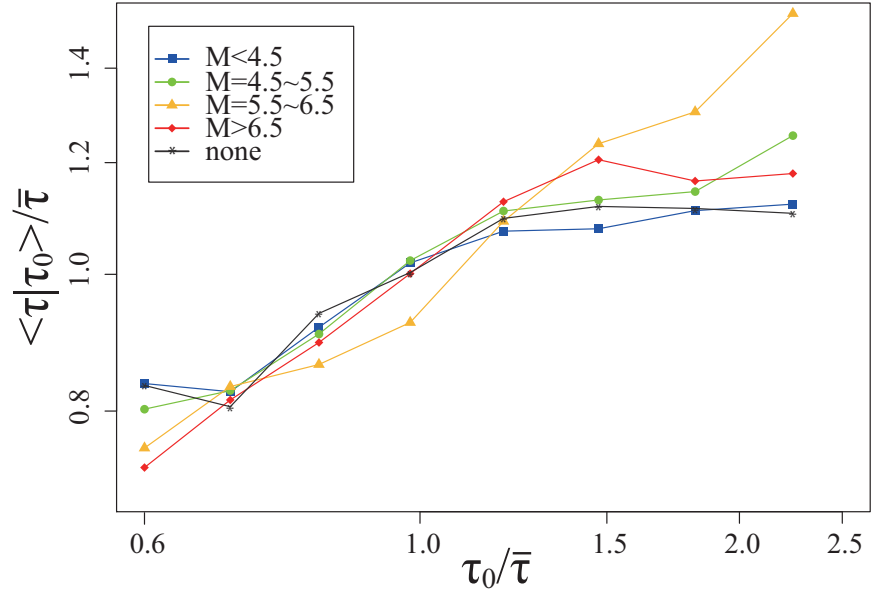}\\
\caption{
 Scaled mean conditional interval $\langle \tau|\tau_0 \rangle / \overline \tau$  vs  $\tau_0 / \overline \tau$ . 
Five groups, one with no proximal earthquake and earthquakes with Richter magnitude 
scale  $M< 4.5 $, $M=4.5\sim 5.5$, $M=6.5\sim 6.5$, $M>6.5$.  An increasing trend 
implies a short-range correlation in the interval series. }
\label{2}

\end{figure}

\begin{figure}[b]
\centering \includegraphics[width=0.6\textwidth]{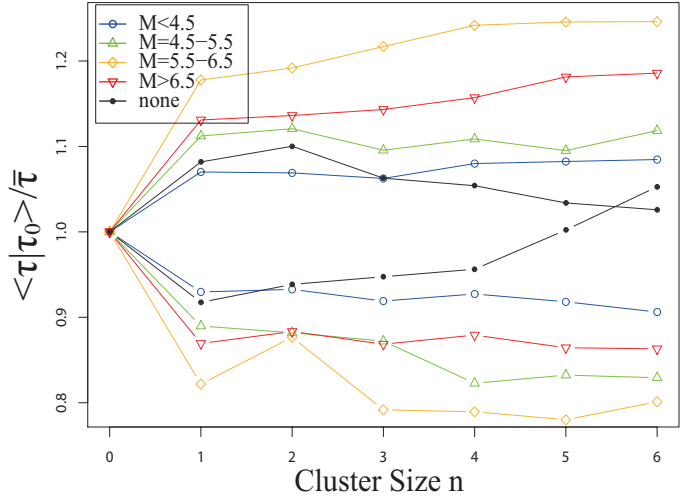}\\
\caption{
Long-range memory in interval clusters. $\tau_0$ 
signifies a cluster of intervals, consisting of $n$ 
consecutive values that all are above (denote as "$+$") or 
below (denote as "$-$") the median of the entire interval 
records. Plots display the scaled mean interval conditioned on a
cluster, $\langle \tau|\tau_0 \rangle / \overline \tau$  vs
the size $n$  of the cluster for five group intervals.
The upper part (overplotted) of curves is for "$+$" 
clusters while the lower part is for "$-$" clusters.
The plots show that "$+$" clusters are likely to be
followed by large intervals and "$-$" clusters by
small intervals, consistent with long-term correlations 
in interval records. Similar to Fig.2, the long-term 
correlation increases with earthquake size, with
exceptions for very large earthquakes.
}
\label{3}
\end{figure}

\begin{figure}[b]
\centering \includegraphics[width=0.7\textwidth]{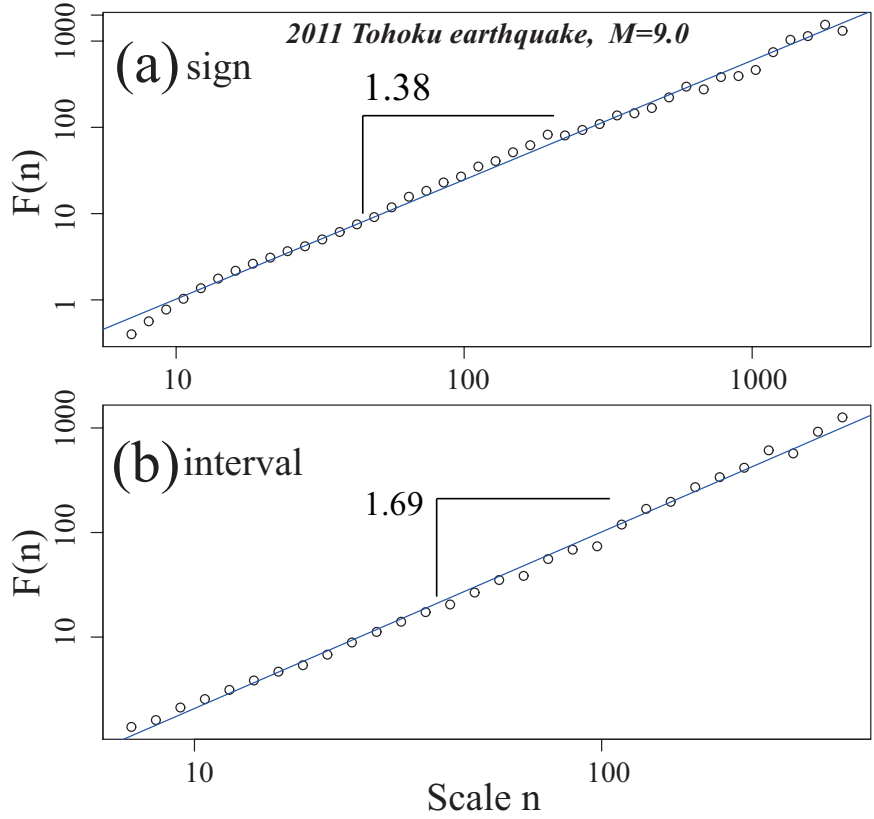}\\
\caption{
DFA fluctuation function $F(n)$ of 2011 Tohoku earthquake  as a 
function of time scale $n$  
 ($ F(n) \propto n^\alpha$)
for (a) sign time series ($\alpha+1=1.32$, ($\alpha<0.5$), 
indicates anticorrelations) and (b) interval time series 
 ($\alpha+1=1.69 $, ($\alpha>0.5$),  indicates correlations).
}

\label{4}
\end{figure}
\begin{figure}[b]
\centering \includegraphics[width=0.6\textwidth]{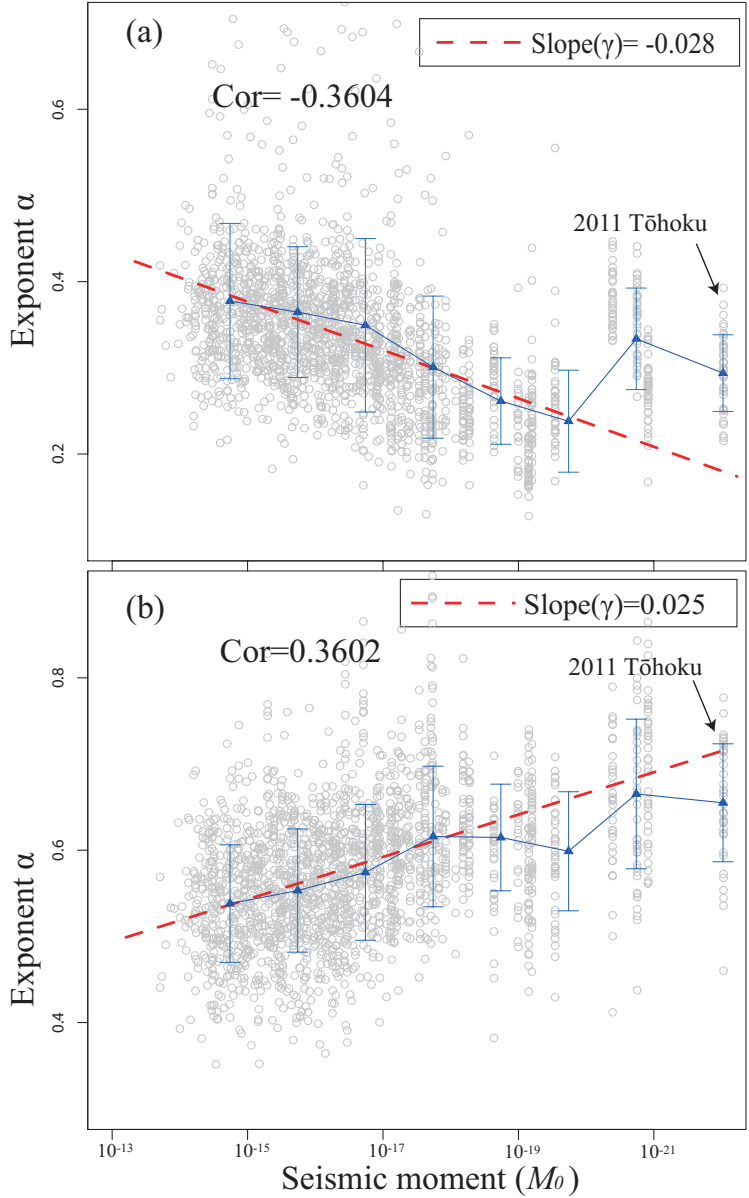}\\
\caption{
 Scaling exponent $\alpha$  vs seismic moment (Richter magnitude scale)
  for (a) sign time series (correlation 
 coefficient $Cor= -0.3604$), and (b) interval time series 
 (correlation coefficient $Cor = 0.3602$).
The values of $\gamma$ show negative slope in the regression 
$\alpha$  vs seismic moment 
 of the sign series, and   positive slope in the regression 
  of the interval series.
  Triangular symbols show the mean of exponent  within each bin
   ( bins:  $<1e+15,1e+15\sim1e+16,1e+16\sim1e+17,1e+17\sim1e+18,1e+18\sim1e+19,1e+19\sim1e+20,>1e+21$),
    the error bar shows the $\pm$ standard deviation.  
    The plots show a linear relationship 
    between logarithmic earthquake moment and scaling exponent
     $\alpha$ in the  sign and interval series,
      with exceptions for
very large earthquakes.}

\label{5}

\end{figure}

\begin{figure}[b]

\centering \includegraphics[width=0.9\textwidth]{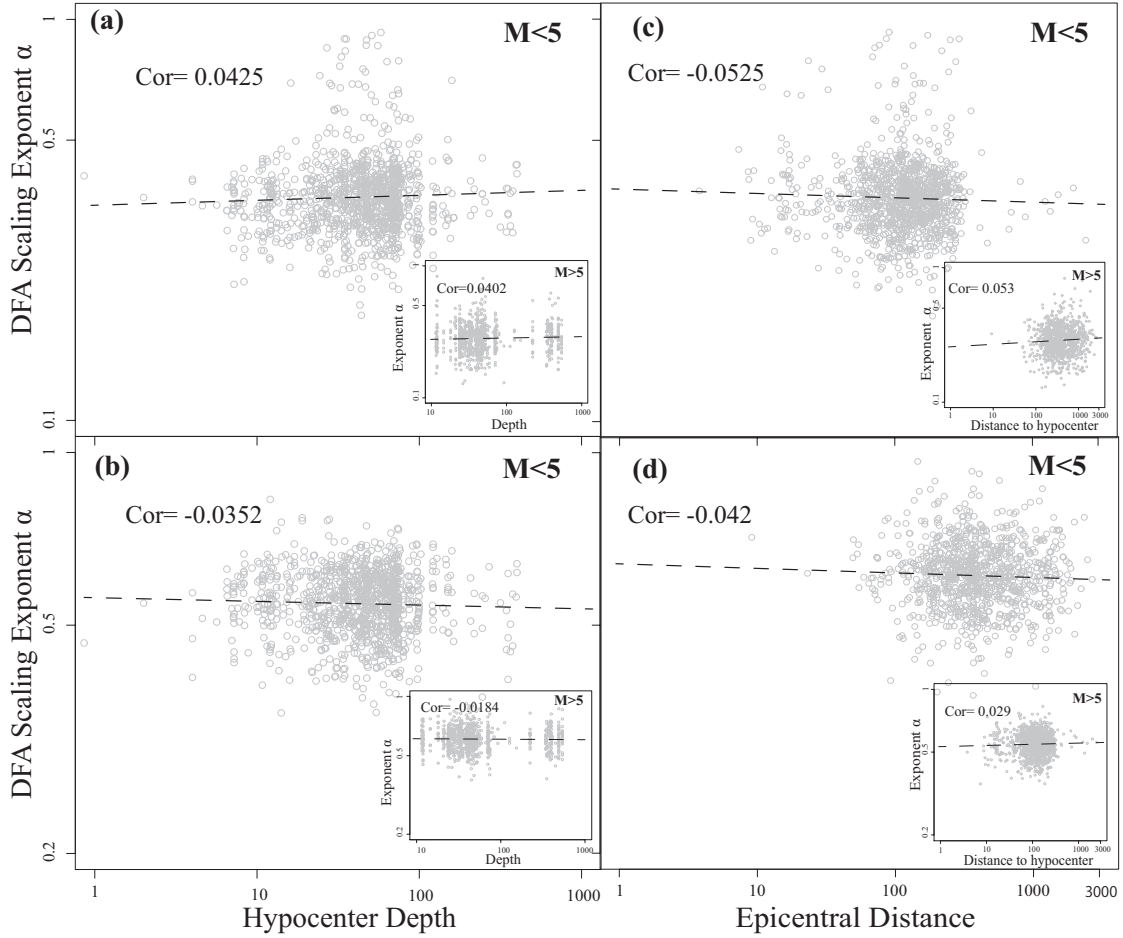}\\

\caption{ Scaling exponent $\alpha$ vs hypocenter depth for events where 
Richter magnitude scale $M<5$ for 
(a) sign time series (b) interval time series. Inset: scaling
 exponent $\alpha$ vs hypocenter depth requiring that Richter magnitude scale $M>5$. 
 (c) and (d) show Scaling exponent $\alpha$ vs epicentral distance for events where 
Richter magnitude scale $M<5$. Inset: scaling exponent $\alpha$ vs epicentral distance requiring that Richter magnitude scale $M>5$. (c) sign time series, (d) interval series.
All absolute values of correlation coefficient are smaller than $0.1$, 
showing that $\alpha$ is uncorrelated with both hypocenter depth 
and epicentral distance.}
\label{6}
\end{figure}
\end{document}